# Numerical simulation of the interactions of domain walls with breathers in two-dimensional O(3) nonlinear sigma model

F. Sh. Shokirov

S. U. Umarov Physical-Technical Institute of Academy of Sciences
of the Republic of Tajikistan, Aini Avenue 299/1, Dushanbe

**Abstract**. By methods of numerical simulation the interaction of 180-degree domain walls with oscillating solitons in (2+1)-dimensional O(3) nonlinear sigma model is investigated. A model of the collisions in which there is observed a reflection of solitons from each other, long-range interactions, as well as the decay of the oscillating soliton are obtained.

**I. Introduction**

This paper presents the results of numerical simulation of the interaction of dynamic (oscillating solitons, breathers) and topological (domain walls, kink/antikink) solitons in the (2+1)-dimensional anisotropic O(3) nonlinear sigma model (NSM).

The Lagrangian and Hamiltonian of the O(3) NSM in isospin parameterization are as follows [1-8]:

$$\mathcal{L} = \tfrac{1}{2}[\partial_\mu s_a \partial^\mu s_a + (s_3^2 - 1)], \qquad (1)$$

$$\mathcal{H} = \tfrac{1}{2}[(\partial_0 s_a)^2 + (\partial_k s_a)^2 + (1 - s_3^2)],$$

$$\mu = 0,1,2; \quad a = 1,2,3; \quad s_a s_a = 1; \quad k = 1,2.$$

Lagrange-Euler equations of the model (1) in the Euler parameterization can be written as:

$$2\partial_\mu \partial^\mu \theta + \sin 2\theta \left(1 - \partial_\mu \varphi \partial^\mu \varphi\right) = 0,$$

$$2\cos\theta \, \partial_\mu \varphi \partial^\mu \varphi + \sin\theta \, \partial_\mu \partial^\mu \varphi = 0,$$

$$\mu = 0,1,2,$$

where $\theta(x,y,t)$ и $\varphi(x,y,t)$ – are Euler angles, associated with isospin parameters of the model (1) as follows [6,8]:

$$s_1 = \sin\theta \cos\varphi, \qquad s_2 = \sin\theta \sin\varphi, \qquad s_3 = \cos\theta;$$

$$s_i s_i = 1, \qquad i = 1,2,3.$$

The fields $s_1$, $s_2$, $s_3$ in this case are the coordinates of the unit isovector $S(s_1, s_2, s_3)$, which describes the dynamics of solutions in isotopic space of the sphere $S^2$. Note also that those equations (1) in the meridian intersection of the sphere $S^2$ ($\varphi(x,y,t) = 0$) are reduced [2-9] to the (2+1)-dimensional sine-Gordon equation (SG) in the following form:

$$2\Box\theta = -\sin 2\theta. \qquad (2)$$

*Breather solutions*

In [8] (preprint) the analytical form of trial functions of the equation (2), which evolve to stable in time periodic radially-symmetric solutions has been established. In the terms of isospin fields the obtained solutions have the following form:

$$s_1 = -\frac{2z}{1+z^2}\cos\varphi, \qquad s_2 = -\frac{2z}{1+z^2}\sin\varphi, \qquad s_3 = \frac{1-z^2}{1+z^2}, \qquad (3)$$

$$z(x,y,t) = \frac{\lambda}{\sqrt{1-\lambda^2}}\frac{\sin\varphi}{\cosh(\lambda x)\cosh(\lambda y)},$$

where $z(x,y,t)$ – are complex function, associated with isospin fields (3) and the Euler angles $\theta(x,y,t)$ and $\varphi(x,y,t)$ as follows:

$$z = x + iy = \frac{s_1 + is_2}{1 \pm s_3} = \operatorname{tg}\frac{\theta}{2}e^{i\varphi}. \qquad (4)$$

Parameters $\lambda(t)$ and $\varphi(t)$ in (3) are determined by the equations

$$\lambda_{tt} + \frac{3\lambda}{2(1-\lambda^2)}\lambda_t^2 - \lambda(2-\lambda^2)\varphi_t^2 - \frac{12\sqrt{1-\lambda^2}}{\lambda^2}\left((1-\lambda^2)\operatorname{arcth}(\lambda) - \lambda\right) = 0,$$

$$\varphi_{tt} + \frac{2-\lambda^2}{\lambda(1-\lambda^2)}\lambda_t\varphi_t = 0,$$

obtained in [8]. In this paper, these solutions were named breather (Fig.1).

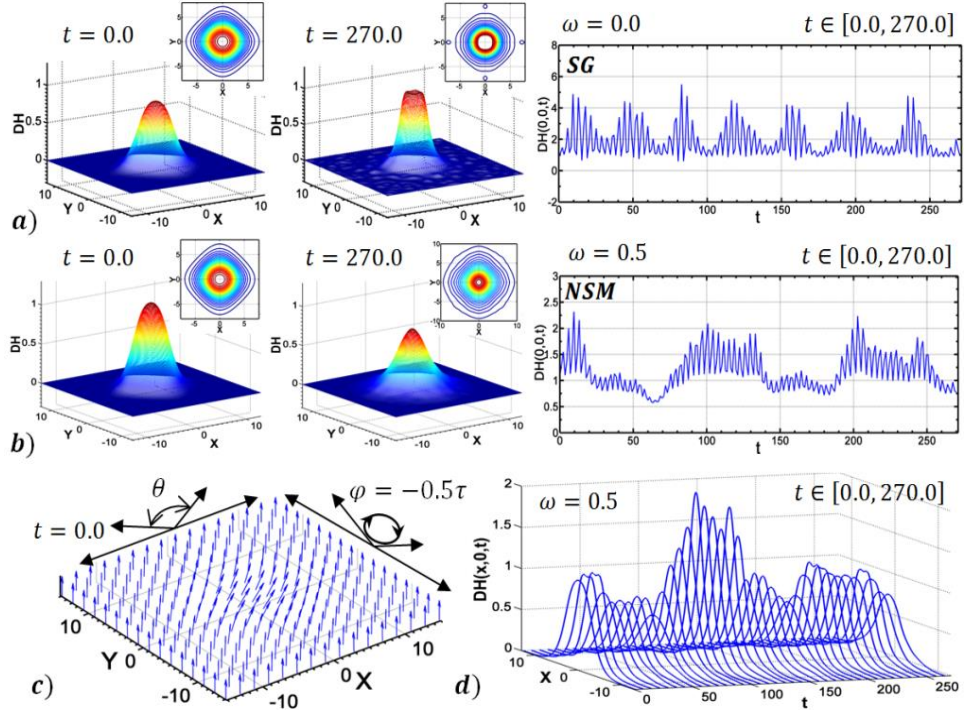

**Fig.1.** Evolution of energy density (DH) of breathers in the O(3) NSM: **a)** breather of the SG equation ($\omega = 0.0$) at $t = 0.0$ and $t = 270.0$, dynamics of point $DH(0,0,t)$; **b)** breather of the O(3) NSM ($\omega = 0.5$) at $t = 0.0$ and $t = 270.0$, dynamics of point $DH(0,0,t)$; **c)** 3D model of isospins of solutions (3) at $t = 0.0$. **d)** Evolution of DH of solutions (3) in planar section $(x, y_0)$, $y_0 = 0.0$. Simulation time: $t \in [0.0, 270.0]$.

*Domain walls*

In [2] on the basis of the given in [9] the analytical form of solution of the Neel type domain walls (DW)

$$z(x,y,t) = 4 arctg\left(e^{B_1\left(\frac{w}{k_1}x - \frac{w}{k_1}x_0\right) + B_2\left(\frac{w}{k_2}y - \frac{w}{k_2}y_0\right)}\right)$$

of the (2+1)-dimensional sine-Gordon equation of the following form:

$$z_{tt} - k_1^2 z_{xx} - k_2^2 z_{yy} + w^2 \sin z = 0$$

was obtained the numerical model of stationary and moving DW of the (2+1)-dimensional O(3) NSM (Fig.2).

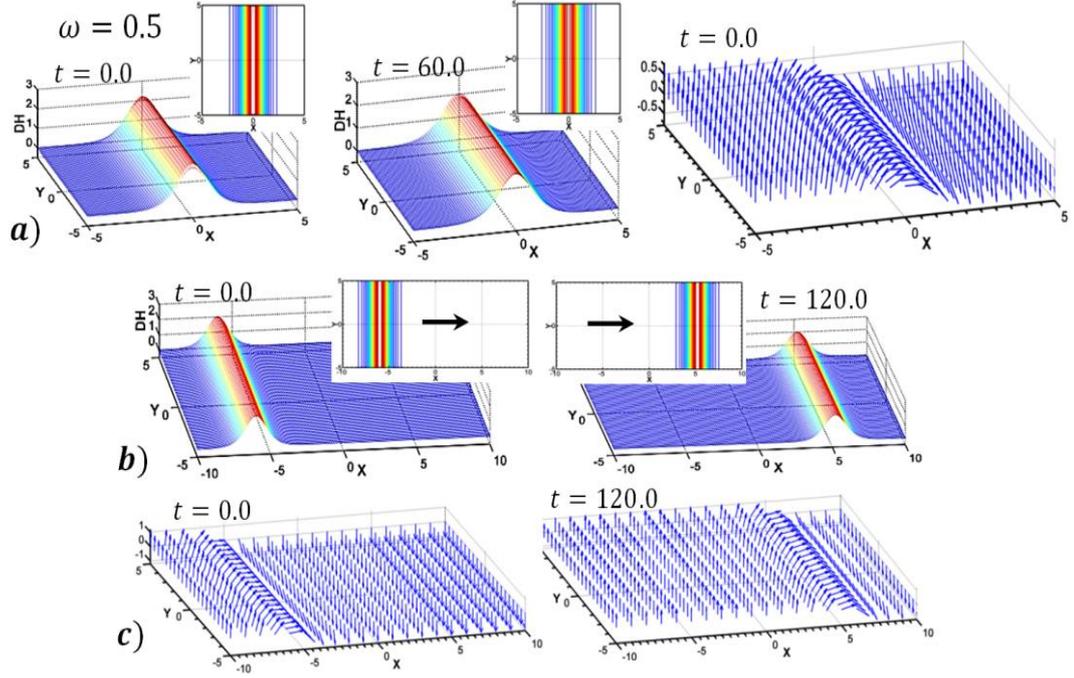

**Fig.2.** Evolution of energy density (DH) of DW (5) in the O(3) NSM ($\omega = 0.5$): **a)** stationary DW at $t = 0.0$ and $t = 60.0$, 3D model of isospins at $t = 0.0$; **b)** moving DW at $t = 0.0$ and $t = 120.0$; **c)** 3D model of isospins of moving DW at $t = 0.0$ and $t = 120.0$. Total simulation time: $t \in [0.0, 120.0]$.

*Interactions*

In present paper, in frame of the (2+1)-dimensional O(3) NSM considered the processes of interaction of breather solutions (3) with DW of type

$$\theta(x,y,t) = 2 arctg\left(e^{B_1\left(\frac{w}{k_1}x - \frac{w}{k_1}x_0\right) + B_2\left(\frac{w}{k_2}y - \frac{w}{k_2}y_0\right)}\right), \quad (5)$$

obtained numerically in work [2].

By methods of numerical simulation are obtained the models of the incident and head-on collisions of solitons (3) and (5), which can be grouped as follows:

1. Interactions of solitons of the SG equation ($\omega_{12} = 0.0$).
2. Interactions of solitons of the O(3) NSM ($\omega_{12} \neq 0.0$).
3. Interactions of DW of the SG equation with breathers of the O(3) NSM ($\omega_1 = 0.0$, $\omega_2 \neq 0.0$).
4. Interactions of DW of the O(3) NSM with breathers of the SG equation ($\omega_1 \neq 0.0$, $\omega_2 = 0.0$).

Numerical experiments of this paper are based on algorithms and difference schemes [10], which developed in works [1,7] for the stationary case, where have been used the properties of the stereographic projection, taking into account the features of theoretical-groups constructions of O(N) NSM class of the field theory (see, e.g., [1-8]).

In the second and third part of the paper presents the results of research into the processes of interactions of one-type solitons – the breathers (3) with DW (5) of the SG equation (2) and O(3) NSM (1) respectively. In all experiments of the second part occurs the decay of the breather at the collision with DW. In this case, is observed the decay of the breather solutions on to the linear perturbation waves. At the collision in the DW appears an oscillatory dynamics – the periodic deformation of energy density (DH), however, the oscillating DW remains stable until the end of simulation time. Note that, as in this case, and in all numerical models obtained in the present paper the DW preserve its stability and integrity of the structure.

In the case of soliton interactions of the O(3) NSM, described in the third part, at all experiments there is observed manifestation of long-range forces. Solitons at approaching to the resonance zone reflected from each other without explicit collision. The second difference of this series of experiments from the previous is that the breather solutions at the (long-range) interaction not destroyed and remain stable. Only in the first experiment of this group are observed a decay of the breather solutions.

In the next two parts of the paper describe the results of the interactions of the different types dynamic and topological solitons of the SG equation and O(3) NSM. The fourth part presents the results of numerical experiments of interaction of DW of the SG equation with breather of the O(3) NSM. In this case, the interaction process is similar to the processes described in the second part, where at the collision occurs a decay of breathers onto the linear radially symmetric waves of perturbation. The only difference between the results of these experiments on the models described in the second part is relatively small amplitude of the deformation of energy concentration (DH) of DW.

In the fifth part of the paper describe the interactions of perturbed ($\omega = 0.5$) DW of the O(3) NSM with breather solutions of the SG equations ($\omega = 0.0$). In this case, there is observed all kinds of interactions described in the preceding subparts – a decay of breather solution, long-range interaction, collision and reflection of the solitons from each other, as well the pass of a localized clot of energy of the breather solution through the DW field. A common feature of all the models obtained in the present paper is to preserve the stability of the DW regardless of the speed of motion of interacting solitons and the type of their collision.

In conclusion, discuss properties of the obtained models and their application in the study of the other tasks.

## II. Interaction of domain walls and breathers of the sine-Gordon equation

In this part of the paper presents the results of numerical simulation of interaction of the domain walls (5) with the breather solutions (3) of the SG equation (2). The experiments were carried out for the solutions of the SG equation in frame of the O(3) NSM at $\varphi(x, y, t) = 0.0$ ($\omega = 0.0$). The following models of the incident and head-on collisions have been obtained:

2.1. $\vec{v}_{DW}(t_0) \approx 0.1 \rightarrow\leftarrow \overleftarrow{v}_{Br}(t_0) = 0.0$;
2.2. $\vec{v}_{DW}(t_0) \approx 0.196 \rightarrow\leftarrow \overleftarrow{v}_{Br}(t_0) = 0.0$;
2.3. $\vec{v}_{DW}(t_0) \approx 0.287 \rightarrow\leftarrow \overleftarrow{v}_{Br}(t_0) = 0.0$;
2.4. $\vec{v}_{DW}(t_0) \approx 0.371 \rightarrow\leftarrow \overleftarrow{v}_{Br}(t_0) = 0.0$;
2.5. $\vec{v}_{DW}(t_0) \approx 0.1 \rightarrow\leftarrow \overleftarrow{v}_{Br}(t_0) = 0.447$;
2.6. $\vec{v}_{DW}(t_0) \approx 0.1 \rightarrow\leftarrow \overleftarrow{v}_{Br}(t_0) = 0.707$.

Speed of motion (at $t_0 = 0.0$) of the interacting solitons in all models of this paper is given by the Lorentz transformation (at $t_0 = 0.0$):

- $\vec{v}_{DW}(t_0)$ – speed of motion of DW (5);
- $\overleftarrow{v}_{Br}(t_0)$ – speed of motion of breather solution (3).

**2.1.** $\vec{v}_{DW}(t_0) \approx 0.1 \rightarrow\leftarrow \overleftarrow{v}_{Br}(t_0) = 0.0$ (Fig.3). At $t = 84.0$ the DW pass a distance equal to $S \approx 8.4$ units, indicating that the full preservation of the initial velocity. It is natural for the solitons of the SG equation, in which unlike from solitons of the O(3) NSM no additional spin dynamics of $\varphi(x, y, t) = 0.0$ in the isotopic space of $S^2$ [1-8].

When approaching to breather (at $t \approx 100.0$) the speed of DW movement gradually decreases and there is observed a certain oscillating movement of its energy density (DH) in DW. In the time interval $120 < t < 160$ there occurs a direct interaction of solitons, which results in the decay of the breather solution. Under the influence of the dynamics of DW the breather completely radiates its energy in the form of linear waves of perturbation (that are absorbed by special boundary conditions on the edges of simulation area).

Note that in all the models derived in this paper at the edges of the field of the numerical integration are set special absorbing boundary conditions [1-8]. After the decay of the breather (at $t \approx 160.0$) the DW with acquired oscillatory dynamics continues to move and completely remains stable until the end of the simulation time – $t = 270.0$. After the decay of the breather solution the speed of DW motion is reduced more than doubled at $t \in [160.0, 270.0]$, and in this time interval it passes a distance equal to $S \approx 5.0$ units.

The energy integral of DW in all experiments of this part preserved with accuracy: $\frac{\Delta En}{En} \approx 10^{-3}$.

Thus, in this experiment, in a frontal collision of DW with breather solution of the SG equation is observed the decay of the dynamic soliton (3) onto the linear waves that propagate along the plane of the DW (and absorbed by boundary conditions on the edges of simulation area). The topological soliton (5) after the collision (at $t \approx 160.0$) remains stable and with the oscillating dynamics continues to move in the original direction until the end of the simulation time $t = 270.0$. In this paper, under the oscillation of energy density (DH) of DW, implies to periodic (temporary) deformation of its energy concentration.

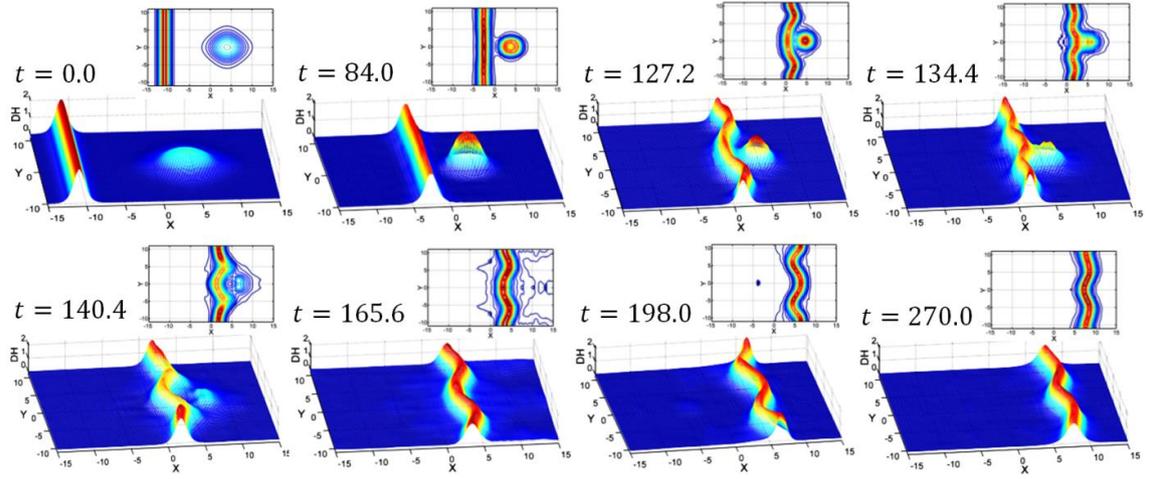

**Fig.3.** Evolution of energy density (DH) of collision process of DW (5), moving at speed $v_{DW}(t_0) \approx 0.1$ with the stationary breather (3) of the SG equation. Simulation time: $t \in [0.0,\ 270.0]$.

The following subparts describe similar experiments, where speed of the incident DW will be gradually increased.

**2.2.** $\vec{v}_{DW}(t_0) \approx 0.196 \rightarrow\leftarrow \tilde{v}_{Br}(t_0) = 0.0$ (Fig.4). At $t = 42.0$ the DW passes a distance equal to $S \approx 8.35$ units, indicating that full preservation of its initial speed, at this stage, as in the previous case. The interaction takes place in analogy to the previous experiment – the breather solution at collision with DW are completely decay onto the linear perturbation waves that propagate along the plane of the DW and absorbed by boundary conditions. The DW is preserved stability and continues to move in the original direction. Here also there is observed a certain strain of energy density (DH) of DW and the occurrence in it the oscillatory dynamics after a collision with a breather.

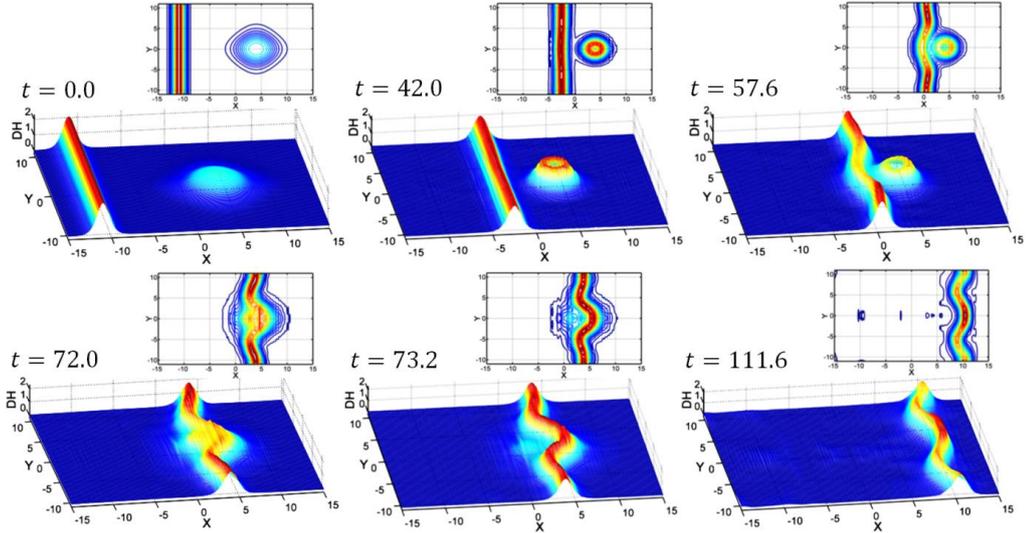

**Fig.4.** Evolution of energy density (DH) of collision process of DW (5), moving at speed $v_{DW}(t_0) \approx 0.196$ with the stationary breather (3) of the SG equation. Simulation time: $t \in [0.0,\ 111.6]$.

**2.3.** $\vec{v}_{DW}(t_0) \approx 0.287 \rightarrow\leftarrow \tilde{v}_{Br}(t_0) = 0.0$ (Fig.5). In this case, as in previous cases, at the interaction is observed a complete decay of breather solution. Fig.5 at $t = 61.2$ and $t = 64.8$ clearly shows the propagation of linear radially symmetric waves formed as a result of decay of the breather. Unlike previous cases, these linear waves propagate not only along the plane of the DW, but also in the direction of movement of the DW and absorbed by the boundary conditions. However, note that as can be seen at $t = 61.2$ and $t = 64.8$ in the resonance zone is stored defined (small) part of the energy density (DH) of breather solution.

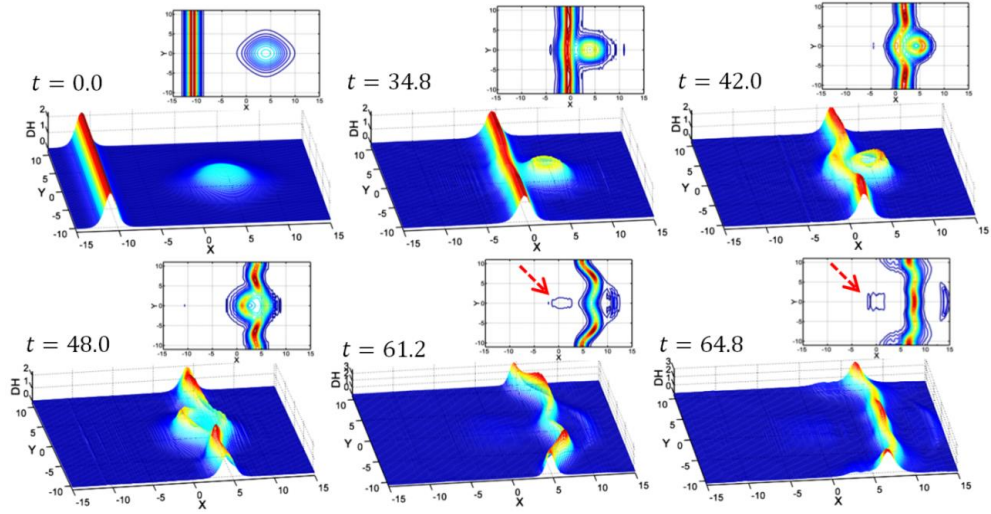

**Fig.5.** Evolution of energy density (DH) of collision process of DW (5), moving at speed $v_{DW}(t_0) \approx 0.287$ with the stationary breather (3) of the SG equation. Simulation time: $t \in [0.0, 64.8]$.

**2.4.** $\vec{v}_{DW}(t_0) \approx 0.371 \rightarrow\leftarrow \vec{v}_{Br}(t_0) = 0.0$ (Fig.6). This model has all the characteristics of the previous models, in addition; in this case more clearly observed a localized region of the energy density (DH) of breather solution, which remained after the interaction of solitons. This fact is clearly seen at $t = 42.0$, $t = 46.8$ and $t = 55.2$

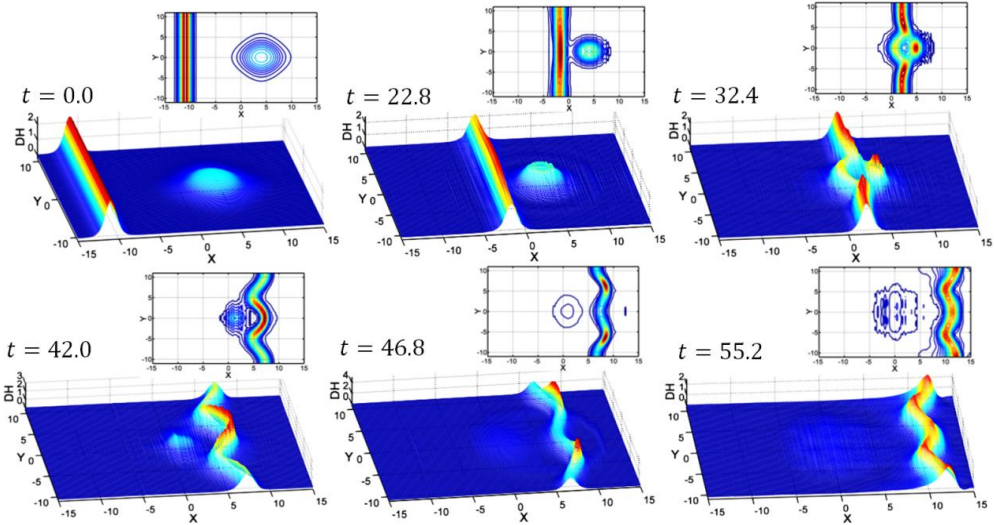

**Fig.6.** Evolution of energy density (DH) of collision process of DW (5), moving at speed $v_{DW}(t_0) \approx 0.371$ with the stationary breather (3) of the SG equation. Simulation time: $t \in [0.0, 55.2]$.

In the following subparts present the results of numerical simulation of head-on collisions of DW and breathers of the SG equation.

**2.5.** $\vec{v}_{DW}(t_0) \approx 0.1 \rightarrow\leftarrow \overleftarrow{v}_{Br}(t_0) = 0.447$ (Fig.7). As in previous experiments, in a collision occurs a complete decay of the breather solution. The DW is similar to the previous cases is oscillating and remains stable in a large enough time interval: $t \in (80.0, 265.0)$. At this stage, the DW continues to move in the original direction and passes a distance equal to $S \approx 6.5$ units.

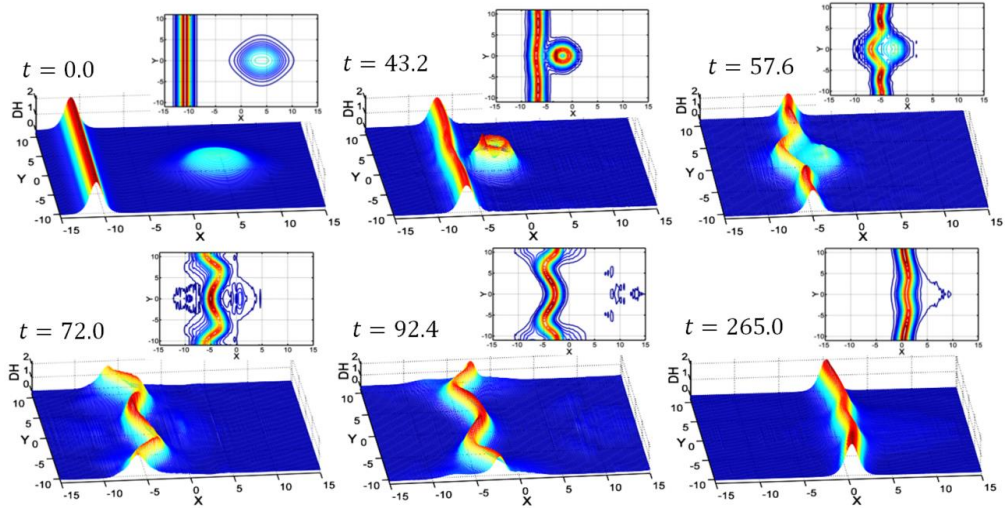

**Fig.7.** Evolution of energy density (DH) of the head-on collision process of DW (5) with breather (3) of the SG equation, moving at speed $v_{DW}(t_0) \approx 0.1$ and $v_{Br}(t_0) \approx 0.447$ respectively. Simulation time: $t \in [0.0, 265.0]$.

**2.6.** $\vec{v}_{DW}(t_0) \approx 0.1 \rightarrow\leftarrow \overleftarrow{v}_{Br}(t_0) = 0.707$ (Fig.8). In this case, after the collision, also is observed the decay of the breather solution, however, part of the energy density (DH) of the breather passes through the DW field. This process is clearly observed at $t = 45.6$, $t = 49.2$ and $t = 50.4$. Next, the linear waves of radiation are absorbed at the boundary of the simulation area; the DW continues to evolve steadily until the end of the simulation time: $t = 254.4$.

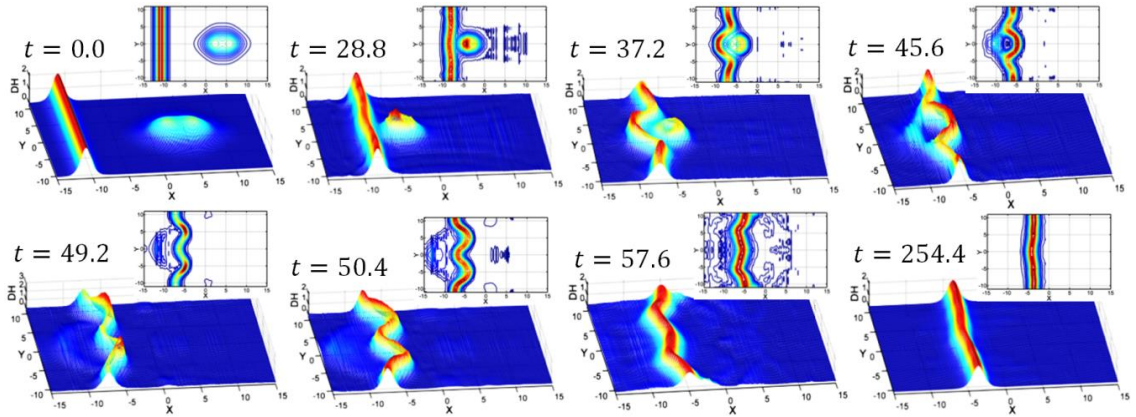

**Fig.8.** Evolution of energy density (DH) of the head-on collision process of DW (5) with breather (3) of the SG equation, moving at speed $v_{DW}(t_0) \approx 0.1$ and $v_{Br}(t_0) \approx 0.707$ respectively. Simulation time: $t \in [0.0, 254.4]$.

In the next section present the results of a similar series of numerical experiments for the O(3) NSM (1).

### III. Interaction of domain walls and breathers in the O(3) NSM

The experiments were carried out for the O(3) NSM at $\varphi(x, y, t) = -0.5\tau$ ($\omega = 0.5$).

**3.1.** $\vec{v}_{DW}(t_0) \approx 0.1 \rightarrow\leftarrow \vec{v}_{Br}(t_0) = 0.0$ (Fig.9). In this case, in the initial stage of the evolution of a system of interacting solitons the breather field gradually fades before the collision process ($t = 79.2$). As will be shown in subsequent experiments, in this case, the energy dissipation of the breather is due to long-range forces.

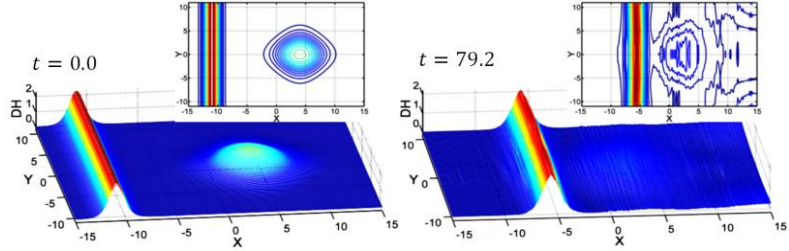

**Fig.9.** Evolution of energy density (DH) of collision process of DW (5), moving at speed $v_{DW}(t_0) \approx 0.1$ with the stationary breather (3) of the O(3) NSM. Simulation time: $t \in [0.0, 79.2]$.

**3.2.** $\vec{v}_{DW}(t_0) \approx 0.196 \rightarrow\leftarrow \vec{v}_{Br}(t_0) = 0.0$ (Fig.10). In this case the long-range effect of interacting solitons form of (3) and (5) has been revealed. Under the influence of the DW field the energy density (DH) of the breather solutions in the time interval $t \in [51.6, 78.0]$ passes a distance equal to $s \approx 4.0$ units in the direction of movement of the DW.

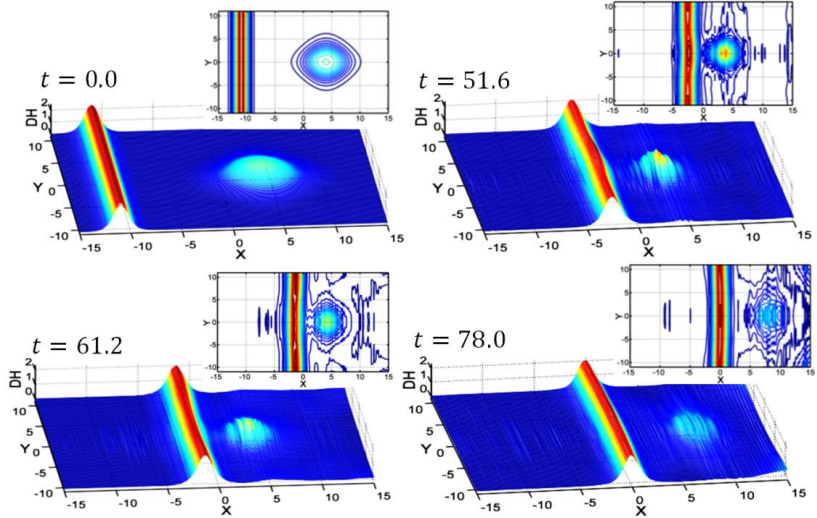

**Fig.10.** Evolution of energy density (DH) of collision process of DW (5), moving at speed $v_{DW}(t_0) \approx 0.196$ with the stationary breather (3) of the O(3) NSM. Simulation time: $t \in [0.0, 78.0]$.

Fig.10 shows that the movement of the breather soliton begins due to the direct interaction of energy densities (DH) solitons (see, e.g., contour projections of DH at $t = 51.6$ and $t = 61.2$). Therefore, in this case and in other similar cases, the long-range effect apparently is relative property of these processes.

**3.3.** $\vec{v}_{DW}(t_0) \approx 0.287 \rightarrow\leftarrow \vec{v}_{Br}(t_0) = 0.0$ (Fig.11). In this case, is also seen the long-range effect (conditional) of the interacting solitons of the O(3) NSM. Under the influence of the DW field the energy density (DH) of the breather solutions in the time interval $t \approx [37.0, 62.4]$ passes a distance equal to $s \approx 8.0$ units in the direction of movement of DW.

Note that unlike the previous part of the experimental results, in part of the O(3) NSM the fluctuations in the energy density of DW at the interaction with the breather solution is not observed.

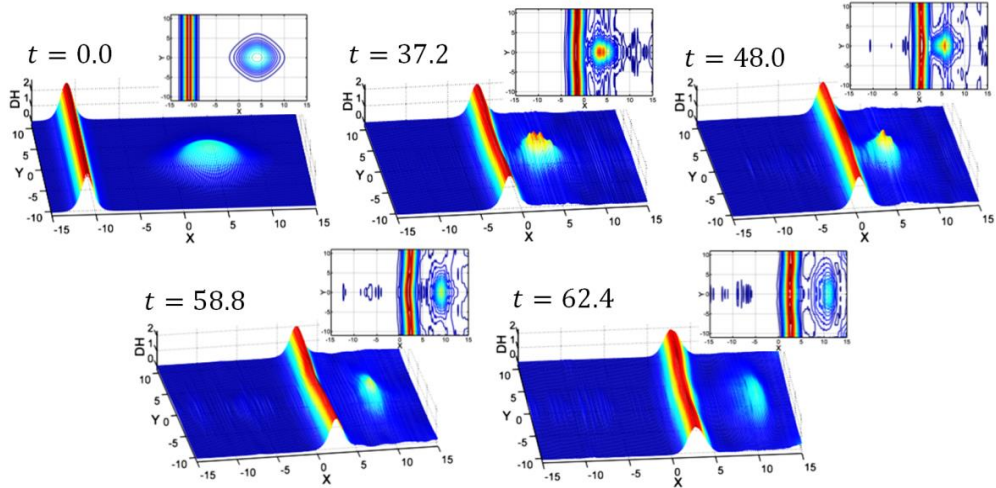

**Fig.11.** Evolution of energy density (DH) of collision process of DW (5), moving at speed $v_{DW}(t_0) \approx 0.287$ with the stationary breather (3) of the O(3) NSM. Simulation time: $t \in [0.0, 62.4]$.

**3.4.** $\vec{v}_{DW}(t_0) \approx 0.371 \rightarrow\leftarrow \vec{v}_{Br}(t_0) = 0.0$ (Fig.12). Similar to previous experiments, in this case also is observed manifestation of the effect of (conditional) long-range interacting solitons of the O(3) NSM. Under the influence of the DW field the energy density (DH) of breather solutions in the time interval $t \approx [25.0, 50.4]$ passes a distance equal to $s \approx 8.3$ units in the direction of movement of DW.

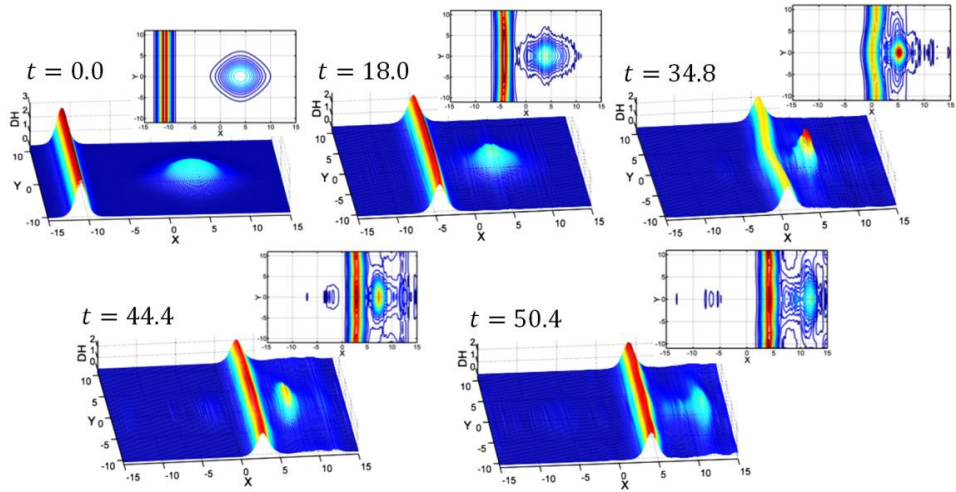

**Fig.12.** Evolution of energy density (DH) of collision process of DW (5), moving at speed $v_{DW}(t_0) \approx 0.371$ with the stationary breather (3) of the O(3) NSM. Simulation time: $t \in [0.0, 50.4]$.

Next, presents the results of similar experiments in a head-on collisions of dynamic (3) and topological (5) solitons of the O(3) HCM (1).

**3.5.** $\vec{v}_{DW}(t_0) \approx 0.1 \rightarrow\leftarrow \vec{v}_{Br}(t_0) = 0.447$ (Fig.13). In this case, as in the previous experiments of this part of paper there is observed a manifestation of long-range forces. At the $t = 50.4$ the DW and the breather passes in opposite directions by a distance equal to $s \approx 4.5$ units. Next, in the time interval $t \in (50.0, 75.0)$ are both soliton moving in an almost standstill and begin to move in the opposite direction. Thus, at $t = 129.6$ solitons is much drifting apart. In this time interval $(75.0 < t < 129.6)$ DW is moved by about one unit ($s \approx 1.0$), while breather solution moves away to a distance equal to $s \approx 7.0$ units. One reason for this difference is in the values of energy integrals of these solitons. In this case, the DW (5) energy is almost twice much energy of breather solution (3):

$$En_{DW}(t_0) \approx 99.32145, \qquad En_{Br}(t_0) \approx 52.77258.$$

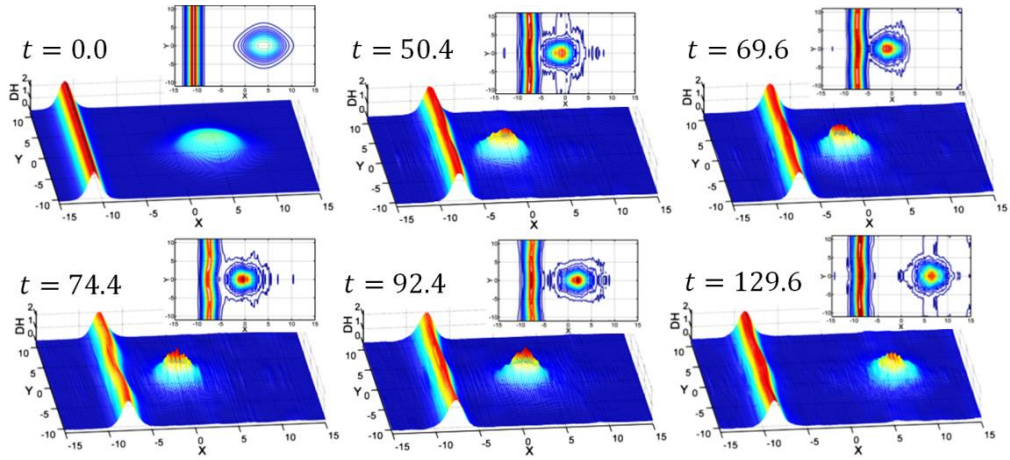

**Fig.13.** Evolution of energy density (DH) of the head-on collision process of DW (5) with breather (3) of the O(3) NSM, moving at speed $v_{DW}(t_0) \approx 0.1$ and $v_{Br}(t_0) \approx 0.447$ respectively. Simulation time: $t \in [0.0, 129.6]$.

In the next subpart present the results of a similar experiment in which the speed of movement of the breather solution increased to $v_{Br}(t_0) \approx 0.707$

**3.6.** $\vec{v}_{DW}(t_0) \approx 0.1 \rightarrow\leftarrow \vec{v}_{Br}(t_0) = 0.707$ (Fig.14). As with previous experiments in this series, in this case also, there is observed a long-range interaction of DW soliton with breather solution.

In this model, the energy of the breather solution is nearly 2/3 from the energy of DW:

$$En_{DW}(t_0) \approx 99.32145, \qquad En_{Br}(t_0) \approx 63.00706.$$

Unlike the previous case, after the collision, the solitons moving away from the resonance zone is about the same distance – $s \approx 4.0$.

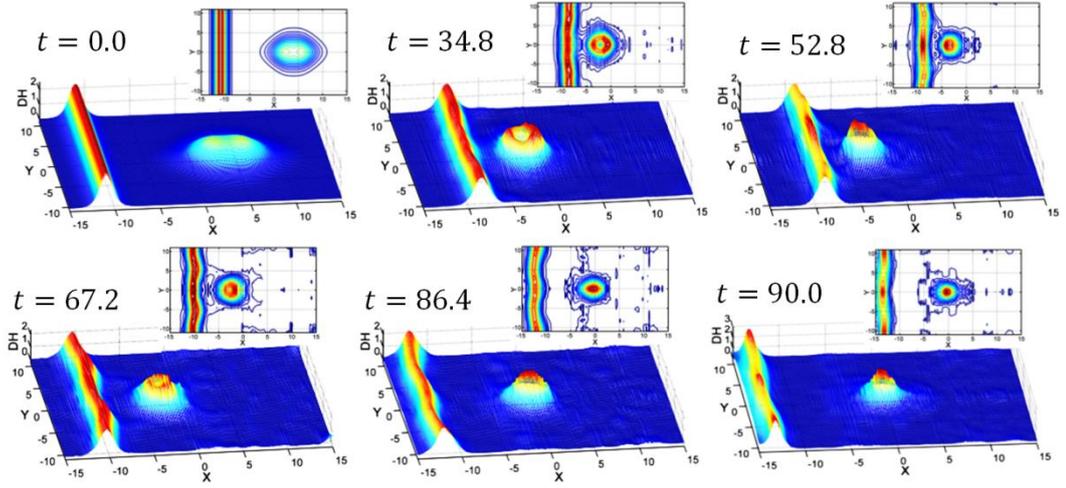

**Fig.14.** Evolution of energy density (DH) of the head-on collision process of DW (5) with breather (3) of the O(3) NSM, moving at speed $v_{DW}(t_0) \approx 0.1$ and $v_{Br}(t_0) \approx 0.707$ respectively. Simulation time: $t \in [0.0, 90.0]$

## IV. Interaction of domain walls of the sine-Gordon equation with breathers of the O(3) NSM

In this part of the paper presents results of experiments for the numerical simulation of the interaction of domain walls (5) of the SG equation (in frame of O(3) NSM at $\omega = 0.0$) with a breather solution (3) of the O(3) NSM in isospin dynamics, which has been added perturbations in form of $\varphi(x, y, t) = -0.5\tau$ ($\omega = 0.5$) [2-8].

**4.1.** $\vec{v}_{DW}(t_0) \approx 0.1 \rightarrow \leftarrow \vec{v}_{Br}(t_0) = 0.0$ (Fig.15). In this case, the breather solution emits almost all its energy under the influence of field of a moving DW.

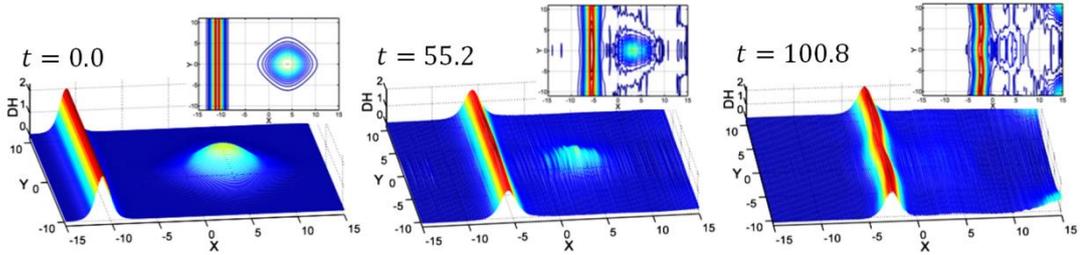

**Fig.15.** Evolution of energy density (DH) of collision process of DW (5) of the SG equation, moving at speed $v_{DW}(t_0) \approx 0.1$ with the stationary breather (3) of the O(3) NSM. Simulation time: $t \in [0.0, 100.8]$

**4.2.** $\vec{v}_{DW}(t_0) \approx 0.196 \rightarrow \leftarrow \vec{v}_{Br}(t_0) = 0.0$ (Fig.16). In this case, similar to the results of the previous experiments, the breather solution is decayed under the influence of field of a moving DW. Unlike the previous case, in this experiment, there is a direct interaction of solitons. After the collision, the DW remains stable and continues to move in the original direction.

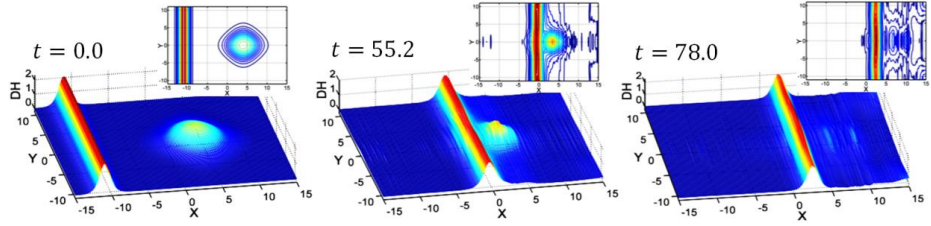

**Fig.16.** Evolution of energy density (DH) of collision process of DW (5) of the SG equation, moving at speed $v_{DW}(t_0) \approx 0.196$ with the stationary breather (3) of the O(3) NSM. Simulation time: $t \in [0.0, 78.0]$.

**4.3.** $\vec{v}_{DW}(t_0) \approx 0.287 \rightarrow \leftarrow \tilde{v}_{Br}(t_0) = 0.0$ (Fig.17). In this case, at the interaction the breather solution (3) decay into linear wave perturbations that spread to the simulation area edges and absorb by the boundary conditions. However, a small fraction of the emitted linear waves, passing the DW field propagates in the direction opposite to the movement of the DW (see, e.g., $50.0 < t < 58.0$). After interacting the DW remains stable and continues to move in the original direction.

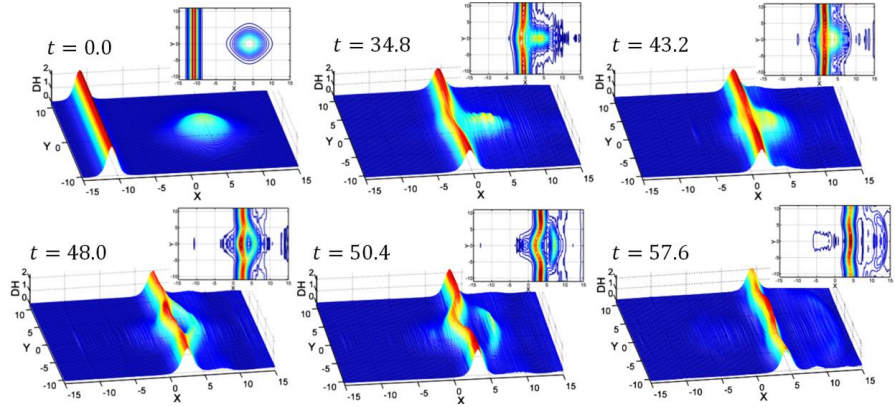

**Fig.17.** Evolution of energy density (DH) of collision process of DW (5) of the SG equation, moving at speed $v_{DW}(t_0) \approx 0.287$ with the stationary breather (3) of the O(3) NSM. Simulation time: $t \in [0.0, 57.6]$.

**4.4.** $\vec{v}_{DW}(t_0) \approx 0.371 \rightarrow \leftarrow \tilde{v}_{Br}(t_0) = 0.0$ (Fig.18). In this case, the solitons interaction process is similar to the previous experimental processes.

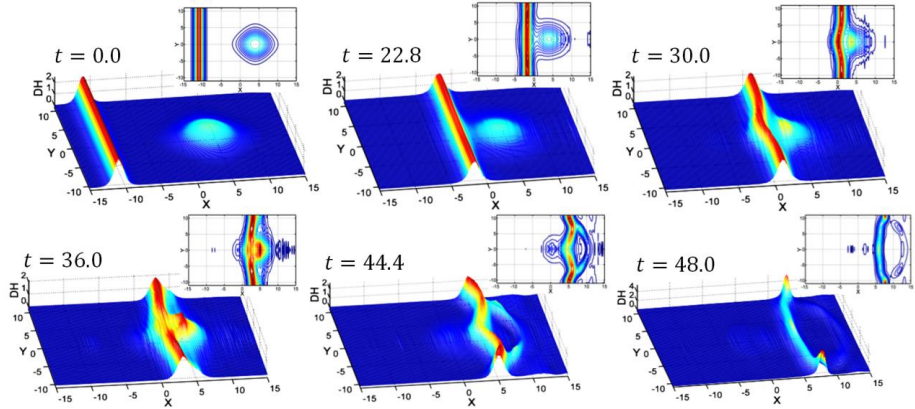

**Fig.18.** Evolution of energy density (DH) of collision process of DW (5) of the SG equation, moving at speed $v_{DW}(t_0) \approx 0.371$ with the stationary breather (3) of the O(3) NSM. Simulation time: $t \in [0.0, 48.0]$.

Next, present the results of experiment of head-on collision of DW of the SG equation with breather solution of the O(3) NSM.

**4.5.** $\vec{v}_{DW}(t_0) \approx 0.1 \rightarrow \leftarrow \overleftarrow{v}_{Br}(t_0) = 0.447$ (Fig.19). In this experiment, at the collision is observed decay of breather solution onto the radially-symmetric linear perturbation waves. Part of these waves passes DW field ($62.4 \leq t \leq 78.0$).

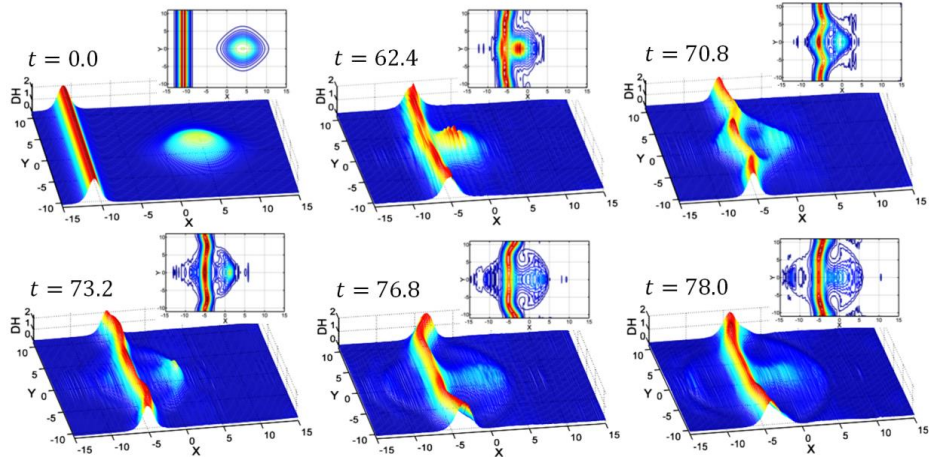

**Fig.19.** Evolution of energy density (DH) of the head-on collision process of DW (5) of the SG equation with breather (3) of the O(3) NSM, moving at speed $v_{DW}(t_0) \approx 0.1$ and $v_{Br}(t_0) \approx 0.447$ respectively. Simulation time: $t \in [0.0, 78.0]$.

Thus, the numerical experiments of this part of the paper have shown that at the interaction of DW of SG equation with breather solutions O(3) NSM, similar to the experiments of the second part, is observed the decay of the breather onto the linear waves. At this the decay of the breather happens by radiation of energy in the form of radially symmetric linear perturbation waves, certain part of which passes DW field.

The following are the results of interaction of DW of the O(3) NSM with breather solutions of the SG equation.

## V. Interaction of domain walls of the O(3) NSM with breathers of the sine-Gordon equation

In this part of the paper presents the results of experiments for the numerical simulation of the interaction of DW (5) of the O(3) NSM ($\omega = 0.5$) with a breather solution (3) of the SG equation (in frame of O(3) NSM at $\omega = 0.0$). In isospin dynamics of the DW field were added a perturbations in the form of $\varphi(x, y, t) = -0.5\tau$ and which moves in the direction of the stationary breather of the SG equation.

**5.1.** $\vec{v}_{DW}(t_0) \approx 0.1 \rightarrow \leftarrow \overleftarrow{v}_{Br}(t_0) = 0.0$ (Fig.20). In this case, the results of experiments similar to those models which have been described in the third part, i.e. in this case, there is also observed a long-range interaction between the DW and breather solution. At the interaction a small portion of the breather energy passes DW field and extends in the opposite direction. The main part of the breather solution under the influence of the field of moving

DW starts moving and reaching the border of simulation area and are absorbed by boundary conditions.

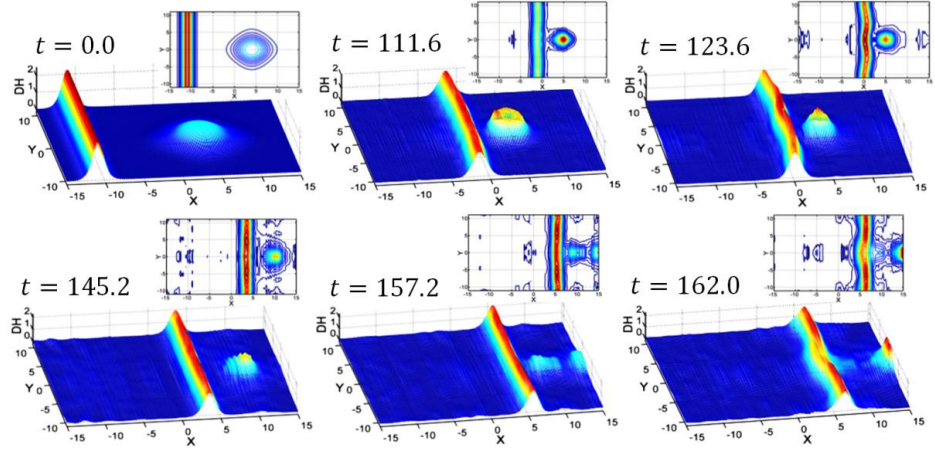

**Fig.20.** Evolution of energy density (DH) of collision process of DW (5) of the O(3) NSM, moving at speed $v_{DW}(t_0) \approx 0.1$ with the stationary breather (3) of the SG equation. Simulation time: $t \in [0.0, 162.0]$.

**5.2.** $\vec{v}_{DW}(t_0) \approx 0.196 \rightarrow\leftarrow \vec{v}_{Br}(t_0) = 0.0$ (Fig.21). In this case, results of experiments similar to the previous case. In contrast to the previous experiments, in this case, at the interaction a relatively large portion of linear waves of perturbations passes through the DW field $(73 < t < 91.2)$.

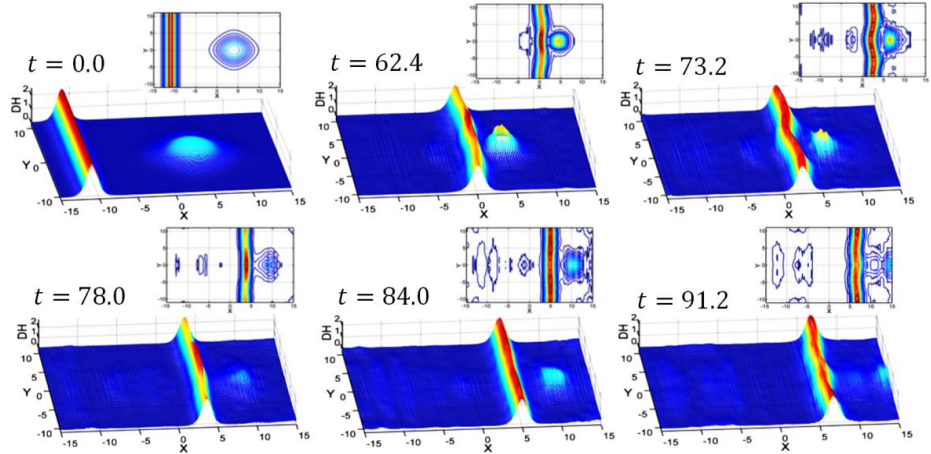

**Fig.21.** Evolution of energy density (DH) of collision process of DW (5) of the O(3) NSM, moving at speed $v_{DW}(t_0) \approx 0.196$ with the stationary breather (3) of the SG equation. Simulation time: $t \in [0.0, 91.2]$.

**5.3.** $\vec{v}_{DW}(t_0) \approx 0.287 \rightarrow\leftarrow \vec{v}_{Br}(t_0) = 0.0$ (Fig.22). In this case, the soliton interaction process is similar to the processes of the previous experiment – part of breather energy at the collision passes through the DW field. The rest part under the influence of moving DW starts to move in the same direction and is absorbed by the boundary conditions. The DW experiencing some perturbations of energy density (DH) and are preserve stability, continues to move in the original direction.

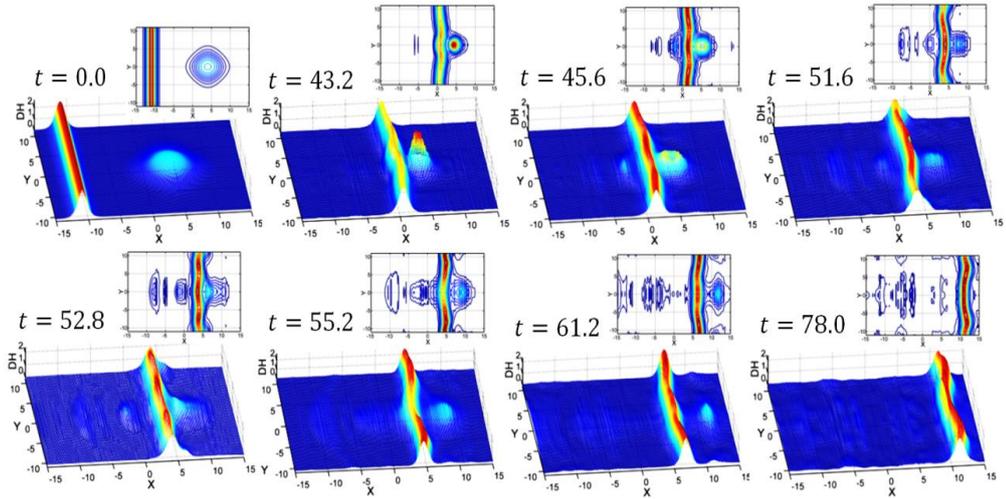

**Fig.22.** Evolution of energy density (DH) of collision process of DW (5) of the O(3) NSM, moving at speed $v_{DW}(t_0) \approx 0.287$ with the stationary breather (3) of the SG equation. Simulation time: $t \in [0.0, 78.0]$

**5.4.** $\vec{v}_{DW}(t_0) \approx 0.371 \rightarrow\leftarrow \overleftarrow{v}_{Br}(t_0) = 0.0$ (Fig.23). In this case, the soliton interaction process is similar to the processes of the previous experiments.

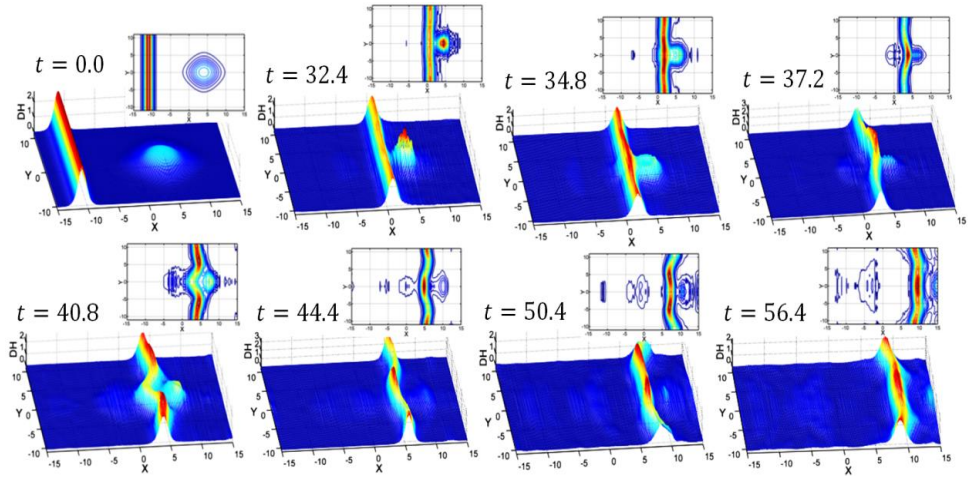

**Fig.23.** Evolution of energy density (DH) of collision process of DW (5) of the O(3) NSM, moving at speed $v_{DW}(t_0) \approx 0.371$ with the stationary breather (3) of the SG equation. Simulation time: $t \in [0.0, 56.4]$.

In next subparts presents the results of a head-on collisions of DW (5) of the O(3) NSM with breather solutions (3) of the SG equation.

**5.5.** $\vec{v}_{DW}(t_0) \approx 0.1 \rightarrow\leftarrow \overleftarrow{v}_{Br}(t_0) = 0.447$ (Fig.24).

In this case, at the interaction of solitons is observed decay of breather soliton onto the linear perturbation waves, certain part of which passes through the DW field.

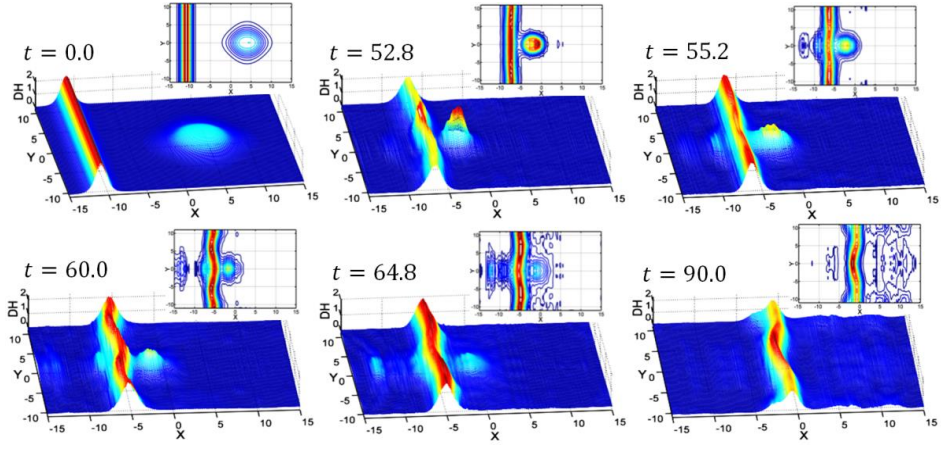

**Fig.24.** Evolution of energy density (DH) of the head-on collision process of DW (5) of the O(3) NSM with breather (3) of the SG equation, moving at speed $v_{DW}(t_0) \approx 0.1$ and $v_{Br}(t_0) \approx 0.447$ respectively. Simulation time: $t \in [0.0, 90.0]$.

**5.6.** $\vec{v}_{DW}(t_0) \approx 0.1 \rightarrow\leftarrow \vec{v}_{Br}(t_0) = 0.707$ (Fig.25). At the collision occurs the decay of the breather solutions by radiation the energy in the form of linear perturbation waves. At the same time as in the previous case, a part of the breather energy passes through the DW field.

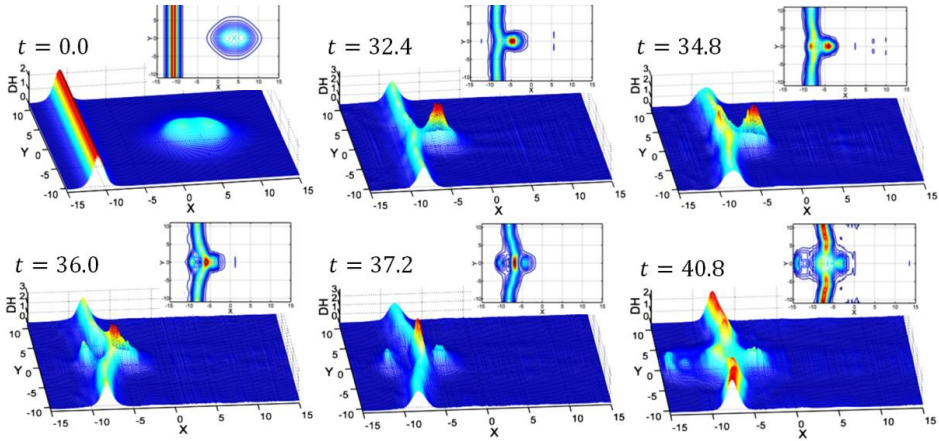

**Fig.25.** Evolution of energy density (DH) of the head-on collision process of DW (5) of the O(3) NSM with breather (3) of the SG equation, moving at speed $v_{DW}(t_0) \approx 0.1$ and $v_{Br}(t_0) \approx 0.707$ respectively. Simulation time: $t \in [0.0, 40.8]$.

Next, presents the results of similar experiments in which the movement speed of DW is increased to $\vec{v}_{DW}(t_0) \approx 0.196$.

**5.7.** $\vec{v}_{DW}(t_0) \approx 0.196 \rightarrow\leftarrow \vec{v}_{Br}(t_0) = 0.447$ (Fig.26). In this experiment also is observed decay of breather solution onto the linear perturbation waves, certain part of which passes through the DW field (see, e.g., $t \in [45.6, 54.0]$).

**5.8.** $\vec{v}_{DW}(t_0) \approx 0.196 \rightarrow\leftarrow \vec{v}_{Br}(t_0) = 0.707$ (Fig.27). In this experiment is observed decay of the breather solutions onto the localized energy clot and linear waves of perturbations. It is clearly seen that at interaction the certain part of DH of breather passes through the DW field (see, e.g., $30.0 < t < 40.0$).

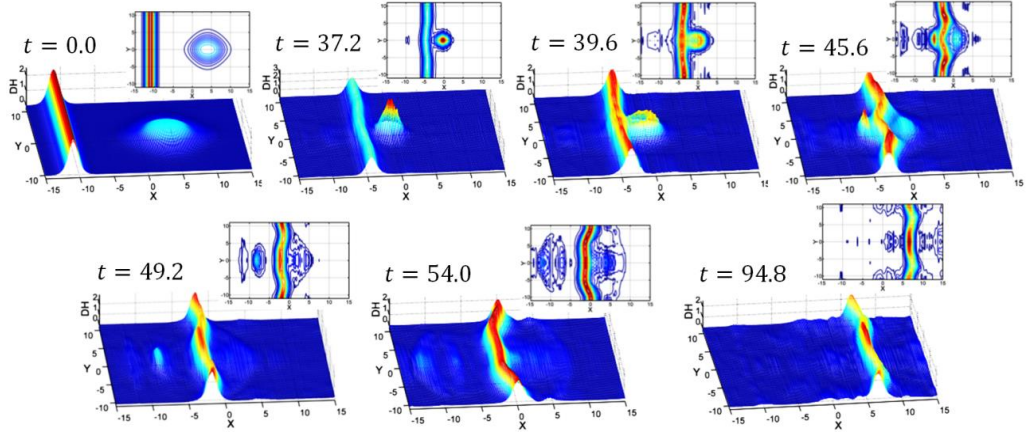

**Fig.26.** Evolution of energy density (DH) of the head-on collision process of DW (5) of the O(3) NSM with breather (3) of the SG equation, moving at speed $v_{DW}(t_0) \approx 0.196$ and $v_{Br}(t_0) \approx 0.447$ respectively. Simulation time: $t \in [0.0, 94.8]$.

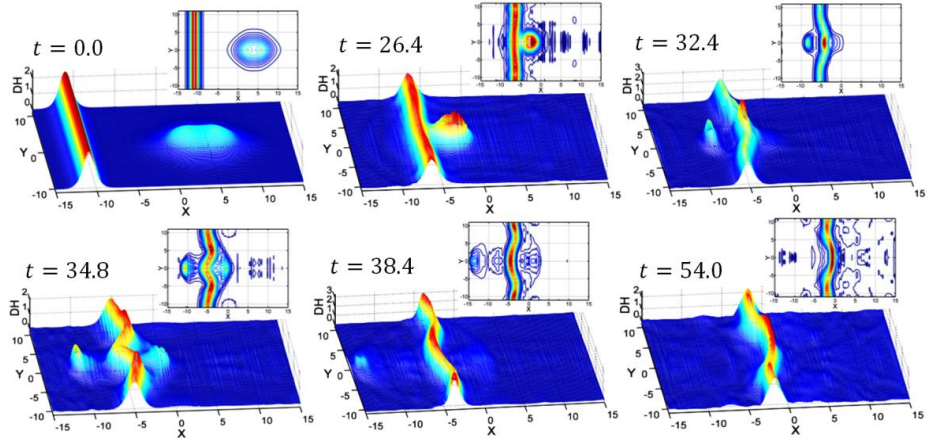

**Fig.27.** Evolution of energy density (DH) of the head-on collision process of DW (5) of the O(3) NSM with breather (3) of the SG equation, moving at speed $v_{DW}(t_0) \approx 0.196$ and $v_{Br}(t_0) \approx 0.707$ respectively. Simulation time: $t \in [0.0, 54.0]$.

Thus, in this part of the paper by a series of numerical experiments is shown that at the interaction of DW (5) of the O(3) NSM with breathers (3) of the SG equation (in frames of O(3) NSM at $\varphi(x, y, t) = 0.0$) the breather solutions decays onto the linear waves of perturbations, certain part of which passes through the DW field. In some cases, there is observed a manifestation of long-range forces (see, e.g., Fig.20 and Fig.21).

The energy integral of DW in all the experiments of the present paper was preserved with accuracy: $\frac{\Delta En}{En} \approx 10^{-3} - 10^{-2}$.

## VI. Conclusion

The numerical results presented in this paper are based on the trial functions for the breather solutions obtained analytically in [8] and on the numerical models of domain walls of the O(3) NSM were obtained in [2].

Note that the DW (5), as might be expected, due to its topological nature are remains stable in all conducted experiments of this paper. The numerical experiments of this paper showed that in contrast to the DW (5) the stability of breather solution (3) in collisions depends, in particular, from the dynamics of isospin field (in space of sphere $S^2$) [8]. In the experiments of second part of paper where the two solitons (3) and (5) evolve without additional dynamics of isotopic spin ($\varphi(x,y,t) = 0.0$) at their interaction occurs the decay of the breather solutions. But as was shown in the third part of the work, at the addition the perturbations ($\varphi(x,y,t) = -0.5\tau$) to the isospin dynamics of both solitons are observed a manifests of a long-range forces between their. In this case both soliton preserve the stability. In all models of this work, where there is direct interaction of solitons, the breather solutions (3) are decayed. The described process is observed in the fourth part and in some experiments of the fifth parts of present paper.

In certain experiments of present work at the direct collision of topological (5) and dynamic (3) solitons a decay of the breather solutions occurs in different ways. For example, in Fig.6 at $t \approx (42.0, 55.2)$ contour projection of the energy density (DH) of system of interacting solitons show that after the collision, a certain part of DH of breather solutions is preserved. In addition, in some experiments in a head-on collision of solitons certain part of energy densities of breather solutions passes through the DW field in the form of well-localized energy clot. For example, in experiments described in fifth part of the work (Fig.24 - Fig.27) clearly observed localized perturbations, which are formed in the backside of the DW (5) at the head-on collisions with breather solutions (3). These localized perturbations are moving in the direction of motion of the breather solutions, before reaching the edges of the simulation area and are absorbed by boundary conditions.

The numerical experiments conducted in this paper can give an overall idea about the nature of the interactions of dynamic and topological solitons of the form (3) and (5) in a two-dimensional anisotropic O(3) NSM [11]. Based on the results of numerical experiments described in this work can be derived theoretical conclusions about the dependence of the interaction properties from the isospin dynamic of localized fields [3].

**Acknowledgments**

The author is very grateful to Prof. Kh. Kh. Muminov for useful discussions.